\documentclass[a4paper,11pt]{article}
\usepackage{pos}
\usepackage{graphicx}

\title{Daya Bay neutrino oscillation progress based on neutron captured on hydrogen}

\author*[1]{Jinjing Li}
\note{For the Daya Bay collaboration}

\affiliation[1]{Tsinghua University,\\
  No. 30, Shuangqing Road, Haidian District, Beijing, PRC}

\emailAdd{lijj16@mails.tsinghua.edu.cn}

\abstract{The Daya Bay reactor neutrino experiment is the first experiment that measured a nonzero value
for the $\theta_{13}$ neutrino mixing angle in 2012. Antineutrinos from six 2.9
$\text{GW}_{\text{th}}$ reactors are detected in eight functionally identical antineutrino detectors deployed in two near (flux-weighted baseline 470 m and 576 m) and one far (1648 m) underground experimental halls. The near-far arrangement of antineutrino detectors allows for a relative measurement by comparing the observed antineutrino rates at various baselines. In 2014, the Daya Bay experiment reported an independent measurement of the nonzero neutrino oscillation parameter $\theta_{13}$, utilizing the data set of neutron captured on hydrogen (nH) with distinct systematic uncertainties from the data set of neutrons captured on gadolinium, and has been improving this measurement since then. The latest result of the Daya Bay nH neutrino oscillation analysis with improved statistics and systematic control is presented.}

\FullConference{%
  40th International Conference on High Energy physics - ICHEP2020\\
  July 28 - August 6, 2020\\
  Prague, Czech Republic (virtual meeting)
}

\begin{document}
\maketitle

\section{Introduction}
Neutrino oscillations are described by the three angles
$\theta_{13}$, $\theta_{23}$, $\theta_{12}$ and CP phase ($\delta$) of the Pontecorvo-MakiNakagawa-Sakata matrix. In particular, the precision of neutrino mixing angle $\theta_{13}$ is of key significance in constraining the leptonic CP phase $\delta$. 
The first measurement of $\theta_{13}$ with a significance greater than five standard deviations was reported by the Daya Bay Reactor Neutrino Experiment in 2012 \cite{Dayabay2012nGd}. In 2014 and 2016, the Daya Bay experiment reported independent measurements \cite{Dayabay2014nH,Dayabay2016nH} of the nonzero neutrino oscillation parameter $\theta_{13}$ using nH data. Recently, we performed a new measurement of electron anti-neutrino oscillation with more than 1958 days of data collection. In this new analysis, we introduced new systematic uncertainty controls and oscillation effect with different neutrino energy ranges is studied.
\begin{figure}[htbp]
\centering
\includegraphics[width=14cm]{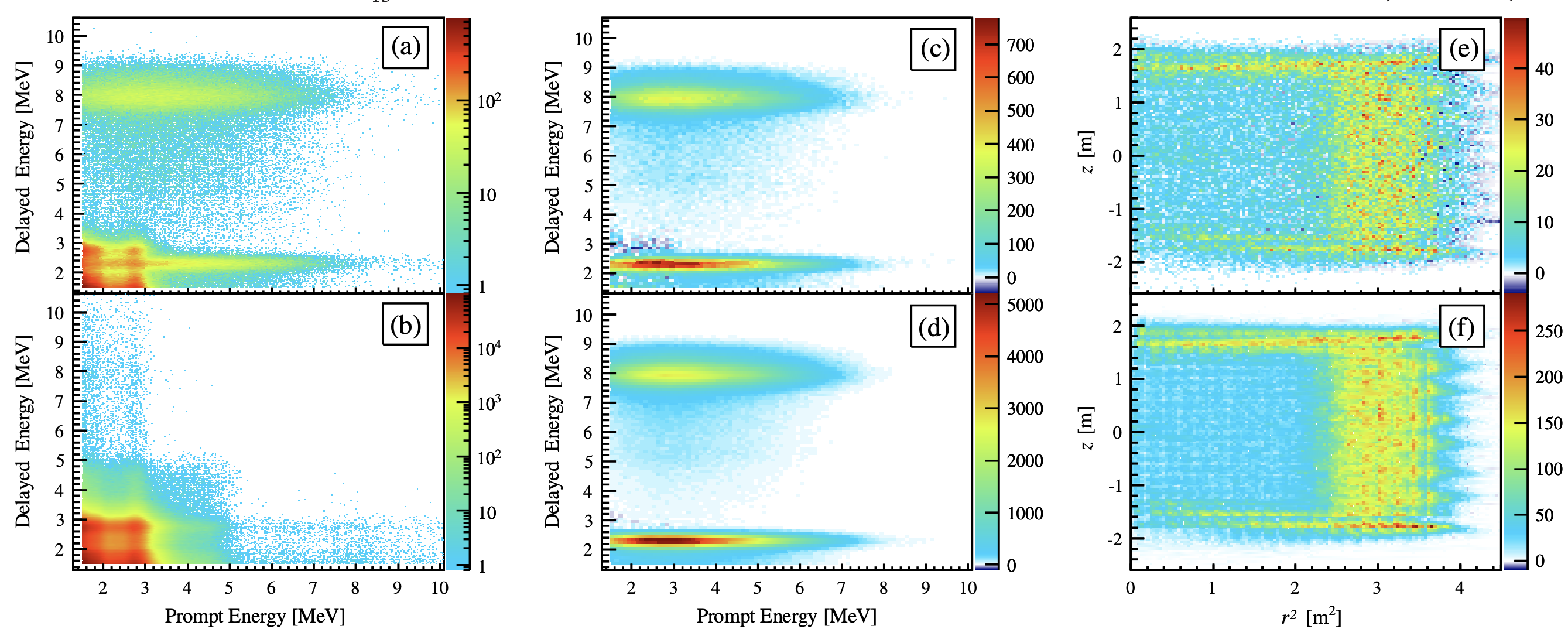}
\caption{This figure is adopted from \cite{Dayabay2016nH} to show the strategy of Daya Bay nH analysis. (a) Distribution of prompt vs. delayed reconstructed energy for all double coincidences with a maximum 50-cm separation in all near-hall ADs, (b) total (621-day) accidental background sample (ABS) for all ADs in the near halls, (c) and (d) are the distributions of prompt vs. delayed reconstructed energy after subtracting the total ABS for the far and near halls, respectively, (e) and (f) are the reconstructed positions of all prompt events after subtracting the total ABS for the far and near halls, respectively.}
\end{figure}
\section{IBD Detection and Background Analysis}
At Dayabay experiment, eight identically designed detectors were deployed in three sites to detect antineutrinos from  six $2.9\text{ GW}_\text{th}$ reactors. The expected number of IBDs in an AD was calculated as the product of the number of IBDs per target proton $\Phi$ and the efficiency-weighted number of target protons $N_\varepsilon$. The calculation of $\Phi$ includes the effect of baseline distance between detector and reactor, the IBD reaction cross section, survival probability of $\bar{\nu}_e$and number of antineutrinos emitted from reactors.
In the three-neutrino-oscillation framework, the survival probability of electron (anti)neutrinos is expressed as
\begin{equation}
    \begin{aligned}
P_{\nu}=& 1-\cos ^{4} \theta_{13} \sin ^{2} 2 \theta_{12} \sin ^{2} \Delta_{21} \\
&-\sin ^{2} 2 \theta_{13} (\cos ^{2} \theta_{12} \sin ^{2} \Delta_{31}+\sin ^{2} \theta_{12} \sin ^{2} \Delta_{32})
\end{aligned}
\end{equation}
where $\Delta_{ij}\equiv1.267\Delta m^2 _{ij}L/E$, $E$ [MeV] is the energy of the neutrino at production, $L$ [m] is the distance between the points of production and interaction of the neutrino, and $\Delta m^2_{ij}$ [eV$^2$] is the difference between the squared masses of mass eigenstates $\nu_i$ and $\nu_j$. 
\par
IBD candidates were selected from pairs of successive events in an AD, including the muon veto, flasher cut, coincidence time cut, distance cut and energy cut etc. Compared to analysis in \cite{Dayabay2016nH}, the basic selection method has been improved. After IBD candidates selection, the resulting prompt-delayed pair events are dominated by IBD events, accidental background and correlated backgrounds. Accidental backgrounds were caused by two uncorrelated AD events that satisfied the IBD selection criteria, and were almost entirely due to natural radioactivity. The accidental rate is calculated based on formulas in \cite{zhe_acc_theory}. And accidental distributions are generated based on the single events collected by each AD. The accidental  distributions then is normalized and subtracted from the IBD candidates, shown in Figure 1. In this new analysis, the estimation of accidental background was improved to reduce the potential bias and uncertainty. While other correlated backgrounds only account for less than $1\%$ of the total background at far site.

\section{Anti-neutrino Oscillation Analysis Result and Progress}
\begin{figure}[h]
\centering
\includegraphics[width=7cm]{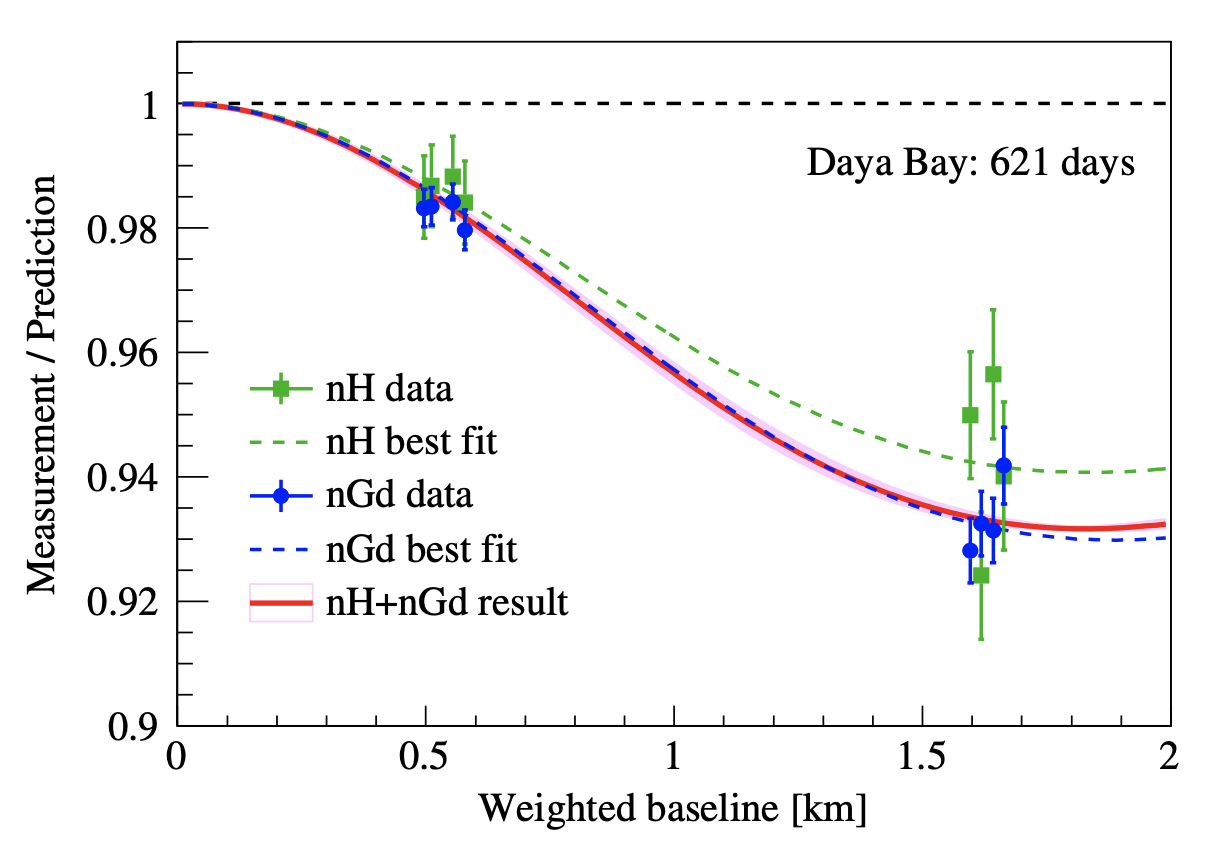}
\caption{Ratio of measured to predicted IBD rate in each detector vs. flux-weighted baseline.}
\end{figure}
Figure 2 shows the $\theta_{13}$ analysis result based on 621 days of data, with $\sin ^{2} 2 \theta_{13}=0.071 \pm 0.011$. In addition to the systematic and statistic improvements in our new analysis, a energy response model is also used for the oscillation analysis with different neutrino energy ranges specified. The energy related effects including energy leakage, energy non-linearity, energy resolution and energy non-uniformity were considered carefully in this model. With all these improvements, a new measurement of $\theta_{13}$  is expected.

\end{document}